\documentclass[9pt,conference]{IEEEtran}
\IEEEoverridecommandlockouts
\usepackage{amsmath,amssymb,amsfonts}
\usepackage{algorithm}
\usepackage{algorithmic}
\usepackage{graphicx}
\usepackage{subcaption}
\usepackage{textcomp}
\usepackage{xcolor}
\usepackage{booktabs}
\usepackage{array}
\usepackage{tikz}
\usetikzlibrary{positioning,fit,backgrounds,arrows.meta,calc}
\usepackage[hidelinks]{hyperref}
\usepackage{needspace}
\usepackage{eso-pic}
\AddToShipoutPictureBG*{\AtPageLowerLeft{\put(\LenToUnit{\dimexpr(\paperwidth-\textwidth)/2\relax},\LenToUnit{0.3in}){%
  \parbox[b]{\textwidth}{\scriptsize\centering \copyright{} 2026 IEEE. Personal use of this material is permitted. Permission from IEEE must be obtained for all other uses, in any current or future media, including reprinting/republishing this material for advertising or promotional purposes, creating new collective works, for resale or redistribution to servers or lists, or reuse of any copyrighted component of this work in other works.}}}}

\renewcommand{\footnoterule}{\kern-3pt\hrule width 0.4\columnwidth\kern2.6pt}

\newcommand{\rc}{R}            %
\newcommand{\cs}{S}           %
\begin{document}

\title{\textbf{Selective Lookahead for Attention-Based Streaming ASR}}

\author{
  \IEEEauthorblockN{Yichen Jia, Bastiaan Tamm, Hugo Van hamme}
  \IEEEauthorblockA{
    \textit{Department of Electrical Engineering (ESAT--PSI)} \\
    \textit{KU Leuven}, Leuven, Belgium \\
    \{yichen.jia, bastiaan.tamm, hugo.vanhamme\}@kuleuven.be
  }
  \thanks{Code: \url{https://github.com/windskylionheart1023/selective-lookahead-asr}}
}

\maketitle

\begin{abstract}
End-to-end attention-based speech recognition is accurate offline but hard to
stream: outputs can depend on future audio, and a little future context per layer
makes the lookahead grow with the number of layers. We address this with two
mechanisms. A \emph{bounded-lookahead chunk encoder} caps every chunk's future
receptive field at a constant number of chunks, independent of the number of layers,
via one age-selection rule shared by self-attention and the depthwise convolution. On
this encoder, \emph{dynamic future-chunk decoding} lets a per-token trigger commit a
token or wait and re-decode it; we propose a learned trigger as the general mechanism,
with a simple confidence threshold as an effective fallback. On full LibriSpeech test-clean the dynamic
system matches the best static-lookahead accuracy ($6.5\%$) at a median latency of
$306$\,ms versus $860$\,ms for one-chunk static lookahead, and a wait budget bounds the
deferral tail below the static baseline's 90th percentile at $0.1$ points more WER.
\end{abstract}


\section{Introduction}

End-to-end automatic speech recognition (ASR) based on the joint connectionist
temporal classification (CTC) and attention architecture~\cite{watanabe2017hybrid} is
accurate offline, where the encoder attends to all frames and the decoder sees the
complete utterance. \emph{Streaming} ASR must emit tokens from partial audio, which
limits each decision's acoustic context and raises the word error rate (WER).
Live captioning, voice assistants, and simultaneous translation need a transcript
within seconds of speech~\cite{shangguan2021dissecting}, so the central engineering
question is how to move along the latency--accuracy trade-off in a controlled way.

We adopt the standard chunked formulation: split the audio into fixed-size chunks
and decode chunk by chunk. While decoding a chunk, a \emph{trigger} may pause
inference to wait for one or more future chunks; with the added acoustic evidence
the model re-decodes the affected tokens more accurately.

\begin{figure}[t]
\centering
\includegraphics[width=0.86\columnwidth]{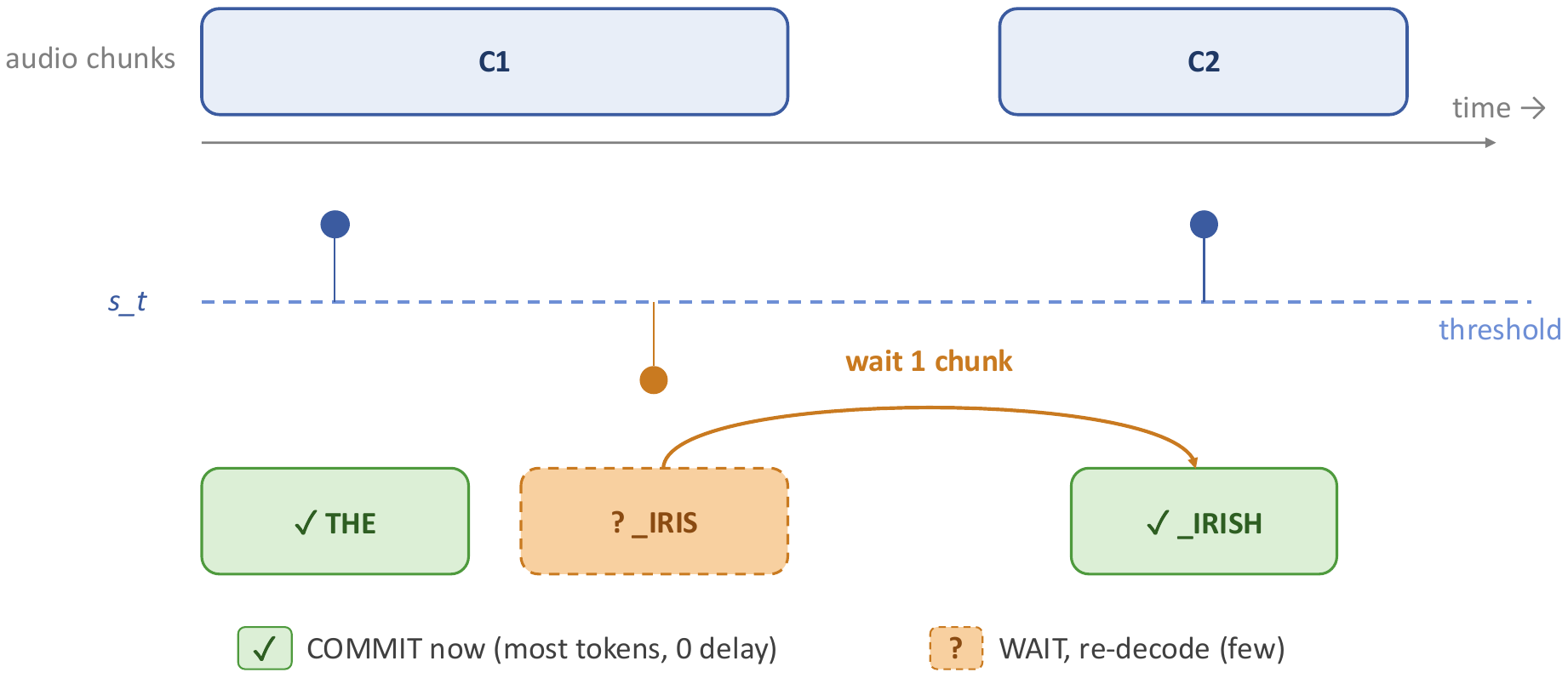}
\caption{Dynamic future-chunk decoding. A per-token trigger commits most tokens
immediately (green, zero added delay) and waits one chunk to re-decode only the
uncertain few (orange): \texttt{\_IRIS} is deferred and resolved to \texttt{\_IRISH}
once the next chunk arrives. $s_t$ is the trigger signal; the dashed line its
threshold.}
\label{fig:teaser}
\end{figure}

Fixed lookahead applies the same delay to every token, and emission
regularizers such as FastEmit~\cite{yu2021fastemit} only change \emph{when} a
fixed-context model emits. We instead decide at inference time, per token, whether
more future acoustic context is worth the delay (e.g.\ committing \texttt{\_IRIS}
early versus waiting one chunk to recover \texttt{\_IRISH}; Fig.~\ref{fig:teaser}).
Two design questions organize the paper:
\begin{itemize}
\item \textbf{Q1 (training/encoding).} How can the model be trained to
\emph{use} future chunks effectively, without its per-chunk receptive field growing with the number of layers?
\item \textbf{Q2 (inference).} What signal reliably indicates \emph{when} waiting
for a future chunk actually helps, so that delay is added only where it improves
accuracy?
\end{itemize}

Our contributions are threefold. 1) We introduce a \emph{bounded-lookahead chunk
encoder} whose self-attention and depthwise convolution share one age-selection rule.
The rule caps each chunk's future receptive field at a tunable constant number of chunks
regardless of the number of layers, and lets a committed chunk be re-encoded under
more context at inference (Section~\ref{sec:encoder}). 2) We propose \emph{dynamic
future-chunk decoding}, a chunk-synchronous beam search whose per-token trigger commits
a token or waits and re-decodes it under richer encoder context
(Section~\ref{sec:decode}). 3) We characterize the chunk-size/lookahead/latency
trade-off on measured per-word emission latency against streaming baselines; the
dynamic trigger defers only where future context changes the token
(Section~\ref{sec:exp}). We will release code and configurations upon acceptance.

\section{Background and Related Work}\label{sec:related}
\textbf{Streaming ASR.} End-to-end ASR rests on three model families, each with a
streaming variant. Connectionist temporal classification
(CTC)~\cite{graves2006ctc,graves2013speech} is frame-synchronous and naturally
streamable. Neural transducers (RNN-T)~\cite{graves2012transduction} add a
label-context predictor and support much on-device streaming
ASR~\cite{rao2017exploring,he2019streaming}, often with a second rescoring
pass~\cite{sainath2019twopass}. Attention-based
encoder--decoders~\cite{chan2016listen,watanabe2017hybrid}, which we adopt for their
offline accuracy, are the hardest to stream because attention is inherently global.
Across families, methods differ mainly in \emph{where} they control latency: the
encoder's streaming receptive field, or the timing of emission.

\textbf{Attention-based streaming.} Most attention-based methods restrict the encoder
to a streaming receptive field: blockwise/chunked
processing~\cite{tsunoo2019contextual,zeineldeen2024chunked},
dynamic-chunk training (U2/WeNet)~\cite{zhang2020unified,yao2021wenet}, a right-context
memory bank (Emformer)~\cite{shi2021emformer}, simulated future
context (CUSIDE)~\cite{an2022cuside}, and bounded per-step right
context~\cite{le2024drc,xia2025mfla}, all with a \emph{fixed} lookahead per token.
Two of these also hold the lookahead independent of the number of layers: Emformer
hard-copies its right-context frames from the input so lookahead cannot leak across
layers~\cite{shi2021emformer}, and the dual causal/non-causal attention of Moritz et
al.~\cite{moritz2021dual} restricts future keys to their causal versions, capping the
lookahead at one layer's window. Our age-versioned encoder differs in three ways. Its
$\rc{+}1$ versions are graded: a future key holds the partial future context it has
already absorbed instead of none. One age rule covers the Conformer convolution as well
as self-attention. And the versions let a committed chunk be \emph{re-encoded} under
richer context, which dynamic decoding relies on (Sections~\ref{sec:encoder},
\ref{sec:decode}).

\textbf{Latency control.} A complementary line regularizes \emph{emission} timing at
training, as in FastEmit~\cite{yu2021fastemit}, TrimTail~\cite{song2023trimtail}, and
delay-penalized transducers~\cite{kang2023delay}, but it only adjusts
emission timing within a fixed acoustic context. Triggered
attention~\cite{moritz2019triggered} fires the decoder from CTC spikes at fixed
lookahead, adaptive non-causal transducers~\cite{strimel2023ancat} learn an
input-adaptive per-frame lookahead inside a single-pass transducer, and monotonic chunkwise
attention~\cite{chiu2018mocha} decides per token when to stop reading input, within a
soft window set at training. Closest to us, asynchronous revision~\cite{huang2020asynchronous}
re-decodes committed history under later right context at a fixed schedule, and the Scout
network~\cite{wang2020scout} learns word boundaries at which to stop reading. We instead defer an already-decoded token and re-decode it
under a re-encoded input, deciding per token at inference whether one more chunk of
latency is worth it (Section~\ref{sec:decode}).

\section{Bounded-Lookahead Chunk Encoder}\label{sec:encoder}
\subsection{The Growing-Receptive-Field Problem}\label{ssec:grow}
\begin{figure}[t]
\centering
\begin{tikzpicture}[scale=0.42,every node/.style={font=\scriptsize}]
  \begin{scope}
    \foreach \l/\name in {0/L1,1/L2,2/L3,3/L4} {
      \node at (-1,\l) {\name};
    }
    \draw[->] (-0.3,-0.7) -- (6.6,-0.7) node[right]{chunks};
    \fill[blue!12] (3,-0.3) -- (1,3.5) -- (5,3.5) -- cycle;
    \foreach \l in {0,1,2,3}{
      \foreach \c in {0,1,2,3,4,5}{
        \draw[gray!50] (\c,\l) circle (0.12);
      }
    }
    \node at (3,4.2) {symmetric: widens};
    \node at (3,-1.25) {$C_t$};
  \end{scope}
  \begin{scope}[xshift=9cm]
    \draw[->] (-0.3,-0.7) -- (6.6,-0.7) node[right]{chunks};
    \fill[green!14] (3,-0.3) -- (4,-0.3) -- (4,3.5) -- (1,3.5) -- cycle;
    \foreach \l in {0,1,2,3}{
      \foreach \c in {0,1,2,3,4,5}{
        \draw[gray!50] (\c,\l) circle (0.12);
      }
    }
    \node at (2.5,4.2) {asymmetric: constant $\rc$};
    \node at (3,-1.25) {$C_t$};
  \end{scope}
\end{tikzpicture}
\caption{Receptive field of the current chunk $C_t$ (filled) across stacked
encoder layers. A symmetric per-layer mask (left) lets the future lookahead grow with
the number of layers. Our asymmetric mask (right, drawn for $\rc{=}1$) caps the future at a constant $\rc$
chunks, the vertical right edge, while past context can still
grow with the number of layers.}
\label{fig:cone}
\end{figure}
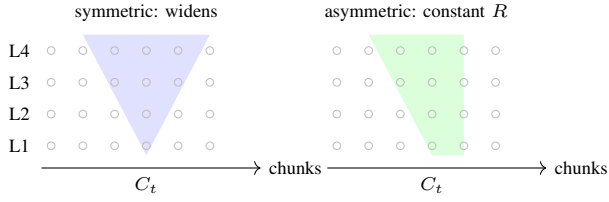

Let the encoder process the input in chunks of $\cs$ frames. Suppose the
``current'' chunk $C_t$ attends to one future chunk $C_{t+1}$ at \emph{every}
layer, and that future chunk in turn looks one step ahead of itself; we call this
the \emph{symmetric} case (Fig.~\ref{fig:cone}). At layer~1, $C_t$ sees
$C_{t+1}$. But at layer~2 the representation of $C_{t+1}$ already absorbed
$C_{t+2}$ at layer~1, so $C_t$ now implicitly sees $C_{t+2}$; by induction, after
$\ell$ layers $C_t$ depends on $C_{t+\ell}$. The effective lookahead grows
linearly with the number of layers~\cite{moritz2021dual}, and the streaming guarantee is lost: the inference engine
would have to wait for far more than one chunk of future audio.
The same compounding arises through
every operation that crosses chunk boundaries: the
Conformer~\cite{gulati2020conformer} convolution module also reads future frames through
its symmetric kernel at every layer~\cite{li2023dynamic}. A streaming bound must cover every cross-chunk path; we
bound attention and convolution in turn, with one shared rule.

\subsection{Attention: Asymmetric Age-Versioned Mask}
We break the recursion by forbidding a future chunk from ever looking further
into the future. Concretely, the current chunk $C_t$ may attend to $L$ past
chunks, itself, and $\rc$ future chunks, while a future chunk attends only to its
own past and itself, never beyond. Because $C_{t+1}$ can never import
$C_{t+2}$ at any layer, $C_t$'s future receptive field is permanently capped at
$\rc$ chunks, \emph{independent of the number of layers}. The mask
is static and identical at every layer.

\begin{figure}[t]
\centering
\begin{tikzpicture}[scale=0.49,every node/.style={font=\scriptsize}]
  \foreach \r in {0,1,2,3}{
    \foreach \i in {0,1,2,3}{
      \draw[gray!55] (\i,3-\r) rectangle (\i+1,4-\r);
    }
  }
  \foreach \r/\cols in {0/{0,1}, 1/{0,1,2}, 2/{0,1,2,3}, 3/{1,2,3}}{
    \foreach \i in \cols {
      \fill[blue!22] (\i,3-\r) rectangle (\i+1,4-\r);
    }
  }
  \foreach \r in {0,1,2,3}{
    \foreach \i in {0,1,2,3}{
      \draw[gray!55] (\i,3-\r) rectangle (\i+1,4-\r);
    }
  }
  \foreach \i/\t in {0/{$t{-}2$},1/{$t{-}1$},2/{$t$},3/{$t{+}1$}} {
    \node at (\i+0.5,4.4) {\t};            %
    \node at (-0.9,3.5-\i) {\t};           %
  }
  \node at (1.8,5.1) {\footnotesize key chunk};
  \node[rotate=90] at (-1.9,2) {\footnotesize query chunk};
  \node[anchor=west] at (4.4,1.5) {\scriptsize current: $L$ past, self, $+1$ future};
  \node[anchor=west] at (4.4,0.5) {\scriptsize future: past $+$ self only};
\end{tikzpicture}
\caption{Chunk-level attention mask for $L{=}2$ past chunks and $\rc{=}1$
(filled $=$ attend). Past rows are drawn in their final versions, which saw their own
$+1$ future when encoded. The current chunk $C_t$ reaches one future chunk; the future
chunk $C_{t+1}$ is available only in its causal version and never reaches beyond
itself, which bounds the receptive field. Rows show only the part of each window inside
this $4{\times}4$ view.}
\label{fig:mask}
\end{figure}
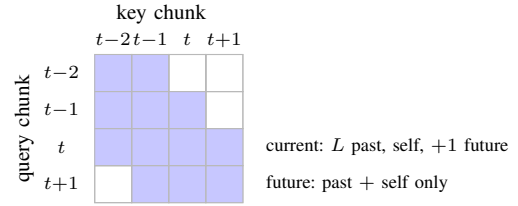

To realize this in a single tensor, we build each chunk in
$\rc{+}1$ \emph{age versions} stacked along one new tensor axis: version $0$ has no future context
(the ``preliminary'' view available the moment the chunk arrives), and each newer
version absorbs one more future chunk. For a query position $q$
in age version $c_q$ and a key position $k$, the readable key version is given by
an age index
\begin{equation}
a(q,c_q,k) = \mathrm{clip}\!\left(\Big\lfloor \tfrac{q}{\cs}\Big\rfloor
- \Big\lfloor \tfrac{k}{\cs}\Big\rfloor + c_q,\; -1,\; \rc\right),
\label{eq:age}
\end{equation}
where $a\!=\!-1$ marks an illegal (too-far-future) key that is masked out;
otherwise the query in version $c_q$ reads key version $c_k\!=\!a$. The upper clip
means that any key chunk far enough in the past is read in its final, fully
contextualized version $\rc$, the only version that must be cached once a chunk is
finalized. A key is also admissible only if it lies within the $L$-chunk past
window of the query (Fig.~\ref{fig:mask}); Eq.~\eqref{eq:age} governs the future side and the version
selection. Age versioning is disabled under full attention (no bound is needed)
and at $\rc\!=\!0$, where the chunked mask is already causal and one version per
chunk suffices.

\subsection{Convolution: Cross-Age Depthwise Convolution}
The convolution path needs the same care as dynamic chunk
convolution~\cite{li2023dynamic}. Making it causal preserves the bound but discards
future audio the attention path may see, while restoring the kernel's natural right
half-window reintroduces the growth of Section~\ref{ssec:grow}, the future horizon
again growing by half a kernel per layer. We instead route the taps by the
\emph{same age-selection rule} as Eq.~\eqref{eq:age}: for the age-$k$ version of chunk
$i$, a tap in chunk $j$ reads age version $\mathrm{clip}(k{-}(j{-}i),0,\rc)$, and taps
with $j{-}i>k$ are zeroed. Every reachable version has a future horizon of at most
chunk $i{+}k$, so the convolution preserves the attention bound at every layer while
still mixing real future audio. Attention and convolution are the only cross-chunk paths, so the bound holds for the
whole network.

\subsection{Efficient Masking and Cost}
The block-structured mask runs on a fast block-sparse kernel,
FlexAttention~\cite{dong2024flex}; the standard fused FlashAttention
kernel~\cite{dao2022flashattention} does not natively express such irregular masks, so
we use it only for the dense fallback. Age-versioning trades compute and memory for
the bound: training builds up to $\rc{+}1$ versions per chunk, and
streaming inference re-encodes each chunk $\rc{+}1$ times as future chunks arrive (for
$\rc\!=\!1$, a preliminary causal pass then a final pass, with past key/value states
cached). This gives the property the dynamic decoder (Section~\ref{sec:decode}) relies
on: because all $\rc{+}1$ versions are seen in training, re-encoding a committed chunk
under more future context is in-distribution.

\needspace{6\baselineskip}
\section{Dynamic Future-Chunk Decoding}\label{sec:decode}
\subsection{Chunk-Synchronous Beam Search with a Trigger}
At inference the encoder emits chunk-causal features and the joint CTC/attention
beam search proceeds chunk by chunk. The CTC prefix
score grows incrementally as each chunk arrives, following the
blockwise-synchronous formulation~\cite{tsunoo2021streaming}; the attention
decoder scores each candidate against the encoder features released so far. After
decoding from the current buffer, a trigger decides whether to \emph{commit} the
new tokens or to \emph{wait} for the next chunk. The model commits tokens up to
the first step at which the trigger fires; from that step on, it waits for the
next chunk and resumes there.

\subsection{Latency Definition}
We measure per-word emission latency: how long after a word's acoustic end the system
commits it. The acoustic end comes from a forced alignment, the Montreal Forced Aligner
(MFA)~\cite{mcauliffe2017montreal}; the commit time is the end of the audio buffer
consumed at commitment. Each word is credited to its slowest-settling sub-word token. Hypothesis words are
matched to reference words by the edit-distance alignment, so latency is measured over
correctly recognized words ($92.7$, $94.0$, and $94.2\%$ of reference words for static
$\rc{=}0$, $\rc{=}1$, and the dynamic system, a similar share in each), and
every evaluation set uses its own MFA reference; wrong words have no anchor and are
excluded, which can censor some long-latency errors. We
report the mean, median, and 90th percentile. Because the acoustic anchor is
system-independent, differences between systems on the same utterances directly measure
added emission delay; each deferred chunk moves the buffer endpoint by one chunk,
$48{\times}\cs$\,ms ($768$\,ms at $\cs\!=\!16$). This is emission delay: decoding runs
faster than real time (Table~\ref{tab:cost}), and we do not add compute time.

\subsection{When Does Waiting Help?}
Two situations motivate waiting. First, a
word whose sub-word pieces span a chunk boundary is prone to a glitch if
committed early. Under limited context, the decoder favors a shorter, locally
probable byte-pair encoding (BPE) piece (e.g.\ \texttt{\_I RI S}) that cannot be
taken back once emitted. Decoded with one more chunk, the decoder recovers the same
word as a single in-vocabulary piece (\texttt{\_IRISH}). Second, tokens whose
identity depends on future context (homophones such as ``write'' vs.\ ``right'')
are ambiguous until later audio arrives.

\subsection{Controller}
The dynamic controller is inference-only. On a defer, every strategy re-encodes the
affected chunk with one more future chunk and exposes the result to both the encoder
and the decoder cross-attention; the strategies differ only in what a defer rolls back
(the last token for A, the whole chunk for B--D) and how it is triggered (Table~\ref{tab:branches}).
A wait budget bounds how long any chunk may defer, and the final chunk always
force-commits. Section~\ref{sec:exp} compares the strategies on \textsc{small}.

\begin{table}[t]
\caption{The four dynamic-controller strategies, differing in what triggers a defer (per token or per chunk) and how much is rolled back and re-decoded (the last token or the whole chunk). Strategy~D, used for all dynamic results, triggers per token and resumes the whole chunk under more context. Rolling back discards the chunk's tokens and re-decodes the chunk from its start; resuming keeps the prefix committed before the trigger fired. Strategy~C defers the chunk when any token's signal fires.}
\label{tab:branches}
\centering
\setlength{\tabcolsep}{6pt}
\begin{tabular}{@{}lll@{}}
\toprule
strategy & trigger & rolls back \\
\midrule
A\ \ token-resume   & per token             & last token \\
B\ \ chunk-rollback & per chunk             & whole chunk \\
C\ \ chunk-rollback & per token, aggregated & whole chunk \\
D\ \ chunk-resume   & per token             & whole chunk \\
\bottomrule
\end{tabular}
\end{table}

\textbf{Commit-stable-prefix.} Whole-chunk rollback (strategies B--D) re-decodes the
entire deferred chunk on each pass and commits it only at the final pass, so every
word in the chunk waits out the full deferral, a heavy latency tail, even
though typically only a short boundary suffix changes across the re-decode passes. We
instead use the local-agreement rule of incremental
decoding~\cite{liu2020lowlatency,polak2022cuni}, committing after each pass the longest
prefix that \emph{agrees} with the previous pass: those words commit as soon as two
consecutive passes agree, while only the uncertain suffix waits for more context. The committed
tokens are unchanged except for the rare word that is locked and would later have flipped
under deeper context: $191$ of $52.6$k hypothesis words on full test-clean, $1.1\%$ of
deferred words, with WER unchanged at $6.5\%$. Accuracy is preserved while the deferral
tail collapses (Table~\ref{tab:dynamic}). The diagnostic curve (Fig.~\ref{fig:frontier})
keeps whole-chunk rollback for comparability. Under strategy~D, deferral without a trigger is not a
distinct baseline: a chunk that always waits its full budget is decoded once, which is
static lookahead; only a selective trigger creates the second pass that local agreement
compares. Algorithm~\ref{alg:defer} states one deferral round.

\begin{algorithm}[t]
\caption{Deferral of chunk $t$ (strategy~D, wait budget $B$)}
\label{alg:defer}
\begin{algorithmic}[1]
\REQUIRE $y$: tokens decoded for chunk $t$ in the current pass; $c$: committed prefix of $y$; $u$: uncommitted rest of $y$
\STATE save beam state $\mathcal{S}$ (hypotheses, decoder and CTC prefix states, encoder buffer); $k \leftarrow 0$; $c \leftarrow$ empty
\LOOP
  \STATE re-encode chunk $t$ with $k$ future chunks; restore $\mathcal{S}$; decode $y$, forced to start with $c$
  \STATE extend $c$ to the longest prefix on which $y$ and the previous pass agree; commit $c$; $u \leftarrow$ rest of $y$
  \STATE \textbf{if} no trigger fires on $u$, \textbf{or} $k = B$ \textbf{then} commit $u$; \textbf{break}
  \STATE $k \leftarrow k{+}1$; wait for the next chunk
\ENDLOOP
\end{algorithmic}
\end{algorithm}

\subsection{Trigger Signals}
A good trigger fires exactly when an early commitment would be wrong. Each trigger reads
a per-token signal $s_t$ and fires against a threshold $\theta$: below $\theta$ for the
confidence family, above $\theta$ for the learned defer probability. We study these two
families.

\textbf{Confidence.} When the model is uncertain about a token, more audio is likely
to help, so a threshold on its confidence is a natural trigger. The raw softmax top-$1$ probability
is poorly calibrated~\cite{li2021confidence,qiu2021learning}, so we evaluate it against two
calibrated alternatives: temperature scaling~\cite{guo2017} and
SR-CEM~\cite{jia2026leveraging}, a lightweight score-rank confidence estimator
for end-to-end ASR that we adapt for the streaming setting. The signal
(top-$1$ probability, top-$1$/top-$2$ margin, entropy, top-$k$ mass) is read on the
decoder step or aggregated over non-blank CTC frames; uncertainty-based emission
control~\cite{sato2026uncertainty} is a related learned alternative.

\textbf{Learned.} Mimicking SR-CEM's setup but skipping the calibrated-confidence
step, we train a \emph{learned trigger} on the same score-rank features to predict the
defer decision directly (whether deferring would fix the token); predicting the
stability of incremental hypotheses this way has
precedent~\cite{selfridge2011stability,mcgraw2012estimating}. Section~\ref{sec:exp}
compares it against the confidence thresholds.

\needspace{6\baselineskip}
\section{Experiments}\label{sec:exp}

We first describe the setup and characterize the chunk-size and lookahead
trade-off, then decompose where streaming errors arise, analyze candidate trigger
signals and their latency cost, and finally evaluate the dynamic controller.

\subsection{Setup}
We use a $12$-layer Conformer~\cite{gulati2020conformer} encoder ($d{=}256$) with
rotary positional encoding~\cite{su2021roformer} and a $6$-layer Transformer decoder,
trained in ESPnet~\cite{watanabe2018espnet} on LibriSpeech train-clean-360 ($\sim$$360$\,h)~\cite{panayotov2015librispeech}
with a $5000$-token unigram vocabulary, jointly with CTC weight $0.3$ (attention
weight $0.7$) and label smoothing $0.1$, optimized with Adam~\cite{kingma2014adam}, Noam
schedule with factor $0.4$ and $25$k warmup steps, batch size $32$, gradient accumulation
$2$, SpecAugment, $40$ epochs, and $10$-best checkpoint averaging by accuracy on
LibriSpeech dev-clean. At inference we use a beam size of $20$, CTC
weight $0.3$, and no external language model. The encoder keeps unlimited left
context, and a deferred token waits at most four future chunks (the training cap). We
vary the confidence threshold and tune the learned trigger on held-out \textsc{dev},
LibriSpeech dev-clean, disjoint from the dev-other evaluation in Table~\ref{tab:splits}. The front end produces one mel frame per
$8$\,ms and the convolutional stem subsamples by a factor of six, so one encoder frame
spans $48$\,ms and a chunk of $\cs$ encoder frames spans $48{\times}\cs$\,ms (e.g.\ $\cs\!=\!16$
is $768$\,ms). Following dynamic chunk training~\cite{zhang2020unified}, we fine-tune
with the chunk size sampled in $[8,32]$
and right context up to four future chunks per utterance, so one model decodes offline
and streams at any chunk size, a dual-mode design~\cite{yu2021dualmode}. Fine-tuning
adds soft early-emission losses (weight $0.1$ each): forced alignments set each token's
deadline chunk, and the decoder cross-attention mass and CTC probability mass falling
after it are penalized, keeping emission timely without a hard cross-attention mask; we evaluate
$\cs\!\in\!\{8,16,24,32\}$ statically (Fig.~\ref{fig:sweep}) and add $\cs\!=\!20$ on the
measured dynamic curve (Fig.~\ref{fig:frontier}). We train every system on train-clean-360, baselines included, so that all results share one
training set; systems trained on the full $960$\,h reach lower absolute WER. Every main accuracy result below is on \emph{full} LibriSpeech test-clean
($2620$ utterances, $52.6$k words), with generalization on test-other and dev-other
(Table~\ref{tab:splits}). The smaller \textsc{small} set ($95$ utterances) is a subset of test-clean and is
harder than the full set: offline $5.5$ vs.\ $4.5\%$. We use it for the chunk-size
trend of Fig.~\ref{fig:sweep}, the fine-tuning gap of Table~\ref{tab:static}, the
controller-strategy comparison, and the cost measurements of Table~\ref{tab:cost}; the
main results are on the full sets.
For out-of-domain transfer (Table~\ref{tab:ood}) we also evaluate $500$-utterance
subsets of \emph{LibriAdapt-US}~\cite{mathur2020libriadapt}, noisy
re-recordings of LibriSpeech; \emph{TED-LIUM~3}~\cite{hernandez2018tedlium}, spontaneous
talks; \emph{WSJCAM0}~\cite{robinson1995wsjcam0}, British read
speech; and \emph{Common Voice US}~\cite{ardila2020common}, crowd-sourced read
speech.
Absolute levels differ across sets. The binomial $95\%$ confidence interval is about
${\pm}0.2$ percentage points on full test-clean ($52.6$k words) and roughly ${\pm}0.8$ on
a $500$-utterance subset, so we read WER differences inside these intervals as ties. We
report paired bootstrap tests over utterances ($10$k resamples) for the main
comparisons. Our WER claims are matched-accuracy comparisons; the gains we report are
in latency.

\subsection{Chunk Size, Lookahead, and the Static Baseline}
\begin{figure}[t]
\centering
\includegraphics[width=0.82\columnwidth]{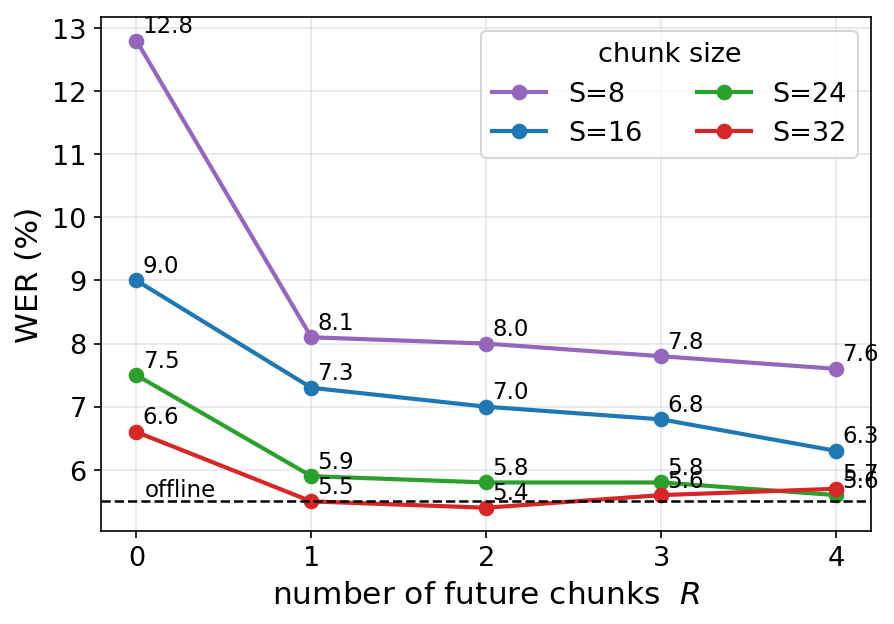}
\caption{Streaming WER vs.\ number of future chunks $\rc$, one curve per chunk
size, on \textsc{small}. These static curves use a chunk-synchronous decoder that
agrees with the dynamic-section decoder (Section~\ref{sec:decode}) to within $0.4$ points at
$\cs{=}16$. Chunk sizes $\cs\!\in\![8,32]$ are in-distribution; the horizontal line is
the offline ceiling ($5.5\%$).}
\label{fig:sweep}
\end{figure}

Fig.~\ref{fig:sweep} shows WER against the number of future chunks $\rc$ for each
chunk size on \textsc{small}. Chunk size is the dominant
factor: enlarging the chunk lowers WER at any fixed $\rc$, with most of the gain
realized by $\cs\!\approx\!32$. Additional lookahead helps most when chunks are
small and flattens for large chunks: a long chunk already contains most of
the locally useful right context. We caution that WER is \emph{not} strictly monotonic in $\rc$ (several points vary by $0.1$--$0.3$ points, within noise), and that chunk sizes outside
$[8,32]$ (e.g.\ $\cs\!=\!48$) lie outside the training range.

We operate at $\cs\!=\!16$, the low-latency end (larger chunks lower WER but proportionally raise latency); there the static streaming WERs on
\textsc{small} are $9.0\%$, $7.3\%$, and $7.0\%$ for $\rc\!=\!0,1,2$, against an
offline reference of $5.5\%$ on the same subset. Table~\ref{tab:static} contrasts it with the un-fine-tuned offline initialization (no chunked FT), which is far worse: $23.8\%$ at $\rc\!=\!0$ and ${\sim}12.5\%$ for $\rc\!\ge\!1$. Chunked fine-tuning is the essential step.

\begin{table}[b]
\caption{Streaming WER (\%, $\downarrow$) at $\cs{=}16$ on \textsc{small}: our
bounded-lookahead chunk encoder vs.\ its un-fine-tuned offline initialization (``no chunked FT'').
Offline is each model's own non-streaming ceiling, not a controlled comparison (the two
differ in training stage); the streaming columns are the contrast.}
\label{tab:static}
\centering
\setlength{\tabcolsep}{3.5pt}
\begin{tabular}{@{}lcccccc@{}}
\toprule
 & offline & $\rc{=}0$ & $\rc{=}1$ & $\rc{=}2$ & $\rc{=}3$ & $\rc{=}4$ \\
\midrule
Chunked FT       & 5.5 & 9.0 & 7.3 & 7.0 & 6.8 & 6.3 \\
No chunked FT    & 6.4 & 23.8 & 12.8 & 12.6 & 12.5 & 12.4 \\
\bottomrule
\end{tabular}
\end{table}

\subsection{Trigger Study}
Deferral only helps errors the decoder makes by committing too early: under limited
context it emits word and phrase duplications at chunk boundaries, which never occur
offline and which waiting for the next chunk can undo. The encoder is a separate limit:
in streaming it has not yet seen future audio, so some tokens stay acoustically ambiguous
no matter how long we wait. Those errors set the gap between the streaming ceiling and offline
accuracy that no trigger can close. We study the two trigger families of Section~\ref{sec:decode} on
held-out \textsc{dev}.

\textbf{Confidence thresholds.} We test three scores as the threshold: the raw softmax
top-$1$ probability, its temperature-scaled~\cite{guo2017} calibration, and the
SR-CEM~\cite{jia2026leveraging} calibrated confidence. All three fall on essentially
one WER--latency curve and reach nearly the same accuracy floor: each ranks the uncertain
tokens similarly enough that the floor is set by the model and chunk size, not by the
confidence estimator. Table~\ref{tab:dynamic} shows the SR-CEM and raw top-$1$
operating points on full test-clean.

\textbf{Learned trigger.} We train a small classifier to predict, for each token,
whether waiting one chunk would turn it from wrong to right; labels come from paired
static decodes of \textsc{dev} at adjacent lookaheads, pooled over $\rc{=}0$--$3$. The
classifier is a $7{\to}32{\to}1$ multilayer perceptron over the seven per-token SR-CEM
features with class-weighted cross-entropy; the features are the token's score, its rank among the beam candidates,
the preceding cumulative hypothesis score, and the four best cumulative candidate
scores at that step~\cite{jia2026leveraging}. We fit it and set its threshold on
held-out \textsc{dev}. On a held-out split of
\textsc{dev}, the confidence score alone orders these
fixable tokens at ROC AUC $0.90$; the learned trigger adds the top-$k$ margins to reach $0.92$,
with the top-1/top-2 margin the most useful of these, while the token's rank among candidates
adds nothing. At the matched $6.5\%$ WER on full test-clean the learned trigger commits
at $630/306$\,ms mean/median versus the SR-CEM threshold's $644/312$ and the raw
top-1's $688/320$ (Table~\ref{tab:dynamic}), a gap of tens of milliseconds; pushed
further, the SR-CEM threshold reaches $6.3\%$ at $919$\,ms mean, and the learned trigger
tracks it at lower thresholds ($6.5\%$ at $\theta{=}0.60$, $731$\,ms mean). We use the learned
trigger for the primary dynamic results because it ranks these tokens best and is the
mechanism we propose. The confidence threshold is a practical fallback: it needs no extra training
and works about as well.

\begin{figure}[t]
\centering
\includegraphics[width=0.75\columnwidth]{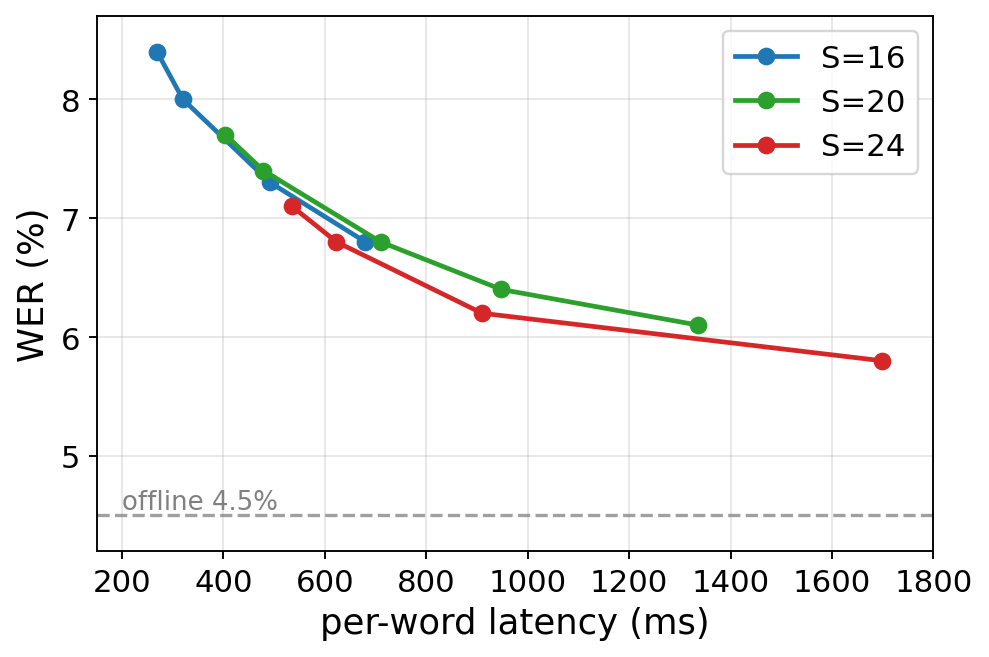}
\caption{Measured per-word mean latency vs.\ WER (strategy~D) at
$\cs\!=\!16/20/24$ on full test-clean. Chunk size is the dominant factor: larger chunks
reach a lower WER floor at higher latency ($\cs\!=\!24$ reaches $5.8\%$), $\cs\!=\!16$
holds the low-latency region, $\cs\!=\!20$ sits between. Latency here is whole-chunk rollback;
commit-stable-prefix (Table~\ref{tab:dynamic}) lowers all three curves.}
\label{fig:frontier}
\end{figure}

\subsection{Dynamic Future-Chunk Results}
Table~\ref{tab:dynamic} reports the $\cs\!=\!16$ latency--WER trade-off on full
test-clean, with per-word emission latency measured against MFA word ends as defined in
Section~\ref{sec:decode}, not a convergence proxy. Among controller strategies we use
strategy~D (chunk-resume in Table~\ref{tab:branches}), a per-token rollback that
re-encodes the whole chunk, fired by the learned wait-policy trigger trained on
\textsc{dev}. On \textsc{small} ($95$ utterances), strategy~D reaches $7.1\%$ WER, the
static $\rc{=}1$ level of the same decoder (the static decoder of
Table~\ref{tab:static} reads $7.3\%$ for this cell); strategies B and C sit within $0.1$
points, inside this subset's noise. The token-only strategy~A trails ($8.0\%$) and
cannot re-encode lookahead, a structural limit, so we use D.

Static lookahead plateaus at $6.5\text{--}6.7\%$ once $\rc\!\ge\!1$ but delays every
token by the same amount: $813$\,ms mean ($860$\,ms median) latency at $\rc\!=\!1$, rising
to $2490$\,ms at $\rc\!=\!4$. The dynamic trigger reaches the best static accuracy
($6.5\%$, static $\rc{=}4$; paired bootstrap $p{=}0.50$) at a \emph{median} latency of
$306$\,ms, a third of static $\rc{=}1$'s $860$\,ms. Its $0.2$-point WER edge over static $\rc{=}1$
is statistically distinguishable under the paired test ($p{=}0.02$) though small in
absolute WER. Two controls on the same pipeline bound what the trigger contributes
(Table~\ref{tab:dynamic}). A \emph{random} trigger that defers the same $34\%$ of tokens
reaches $7.0\%$ at $1078$\,ms median, worse than static $\rc{=}1$ on both axes. An
\emph{oracle} trigger, which reads the reference transcript and defers exactly the chunks
whose first decode contains an error, is the upper bound for any trigger: $6.3\%$ at
$292$\,ms median under whole-chunk rollback, where the learned trigger reaches $6.5\%$ at
$442$\,ms. The tail moves the other way at this operating
point, $p_{90}$ $2800$ versus static $\rc{=}1$'s $1298$\,ms; the budget row below brings
it under. Commit-stable-prefix releases the stable words of a deferred chunk early, so
the deferring minority no longer lengthens the tail against the matched-accuracy anchor:
versus static $\rc{=}4$ the dynamic trigger is lower at every percentile (mean $630$ vs
$2490$, $p_{90}$ $2800$ vs $3518$\,ms; on test-other the dynamic WER is also slightly
lower, $p{<}0.01$). The pattern holds on
\textsc{test-other} and \textsc{dev-other} (Table~\ref{tab:splits}): at a matched
${\sim}17.5\%$ WER the dynamic trigger commits the typical word in ${\sim}463$\,ms
(median) versus static $\rc{=}2$'s ${\sim}1550$\,ms, at higher absolute WER (out of
domain). Varying the trigger across chunk sizes traces the measured WER--latency curve
(Fig.~\ref{fig:frontier}): a larger chunk reaches a lower floor at higher latency, with
$\cs\!=\!20$ a balanced midpoint.

\textbf{Budget bounds the tail.} Capping the wait budget $B$ (Table~\ref{tab:dynamic})
trades a little WER for a much shorter tail: at $\theta{=}0.70$, $B{=}1$ reaches
$964$\,ms $p_{90}$ for $0.3$ points more WER, and it is below static $\rc{=}1$ on mean,
median, and $p_{90}$ at $0.1$ points higher WER ($6.8$ vs $6.7$), inside the test-set
confidence interval. When tail latency matters, $B{=}1$ is the operating point we would
deploy.

\begin{table}[t]
\centering
\setlength{\tabcolsep}{5pt}
\caption{WER and deployment cost at $\cs{=}16$. WER on full test-clean (single stream,
beam $20$); RTF (decoding only, model load excluded) and peak memory on the
$95$-utterance subset, single NVIDIA RTX 2080 Ti.
``waited'' is the fraction of words that use future audio beyond their own chunk (static waits on all, $100\%$).
All configurations run faster than real time. ``Dynamic'' uses strategy~D with the learned wait-policy trigger and commit-stable-prefix (Section~\ref{sec:decode}); its RTF rises modestly with deferral (lower $\theta$).}
\label{tab:cost}
\setlength{\tabcolsep}{4pt}
\begin{tabular}{lcccc}
\toprule
system & WER (\%) & RTF & mem (GB) & waited (\%) \\
\midrule
Static $\rc{=}1$          & $6.7$ & $0.40$ & $3.4$ & 100 \\
Dynamic ($\theta{=}0.70$) & $6.5$ & $0.60$ & $3.5$ & 34 \\
Dynamic ($\theta{=}0.85$) & $7.0$ & $0.51$ & $3.2$ & 19 \\
\bottomrule
\end{tabular}
\end{table}

\textbf{Compute.} All configurations decode faster than real time and use modest
memory (Table~\ref{tab:cost}), so the latency we report is emission delay, not
throughput. With the decoder K/V cache~\cite{vaswani2017attention} enabled, the footprint is set by the encoder,
run by recompute over the left-context window.

\textbf{Out-of-domain datasets.} Beyond LibriSpeech we test $500$-utterance subsets of
four datasets outside the training domain (Table~\ref{tab:ood}), ordered by WER. On noisy
LibriAdapt-US re-recordings the trade-off holds: the dynamic trigger matches the
static WER ($16.3$ vs $16.7\%$, a tie at this sample size, $p{=}0.11$) while committing the typical word in $426$ vs $890$\,ms
(median), a ${\sim}52\%$ latency reduction. Two moderate-WER sets confirm the
gain survives well above LibriSpeech's operating point. On TED-LIUM~3
spontaneous conference speech the dynamic trigger reaches $22.0\%$ at $506$\,ms median,
the one set where deferral also lowers WER significantly ($p{<}0.001$); on WSJCAM0
British read speech it attains static $\rc{=}2$'s $26.0\%$ WER at about a third of its
median latency (Table~\ref{tab:ood}), at the cost of a heavier deferral tail ($p_{90}$
${\sim}3.4$\,s, not shown in the table) that the budget setting bounds. On Common Voice, by contrast, deferral slightly worsens WER
($38.3$ vs $37.5\%$, $p{=}0.05$): the lookahead saturates after one chunk ($\rc{=}1$ and $\rc{=}2$ both
$37.5\%$). Deferral lowers WER on the lower-WER out-of-domain sets but not on Common Voice
at $37.5\%$; as WER rises, more of the errors are intrinsic to the domain and one more chunk
cannot fix them, so future chunks help less. The latency reduction still transfers across
domains, even where the WER benefit does not.

\begin{table}[b]
\caption{Out-of-domain transfer ($\cs{=}16$, $500$-utterance subsets, ordered by WER):
static $\rc{=}1$ and $\rc{=}2$ versus the dynamic trigger (learned wait-policy,
commit-stable-prefix, the same LibriSpeech-dev trigger throughout); latency uses the same
MFA word-end anchoring with per-set reference alignments. Deferral helps up to WSJCAM0's
$26\%$; on Common Voice the lookahead itself saturates ($\rc{=}1$ and $\rc{=}2$ equal)
and deferral no longer lowers WER.}
\label{tab:ood}
\centering
\setlength{\tabcolsep}{3pt}
\begin{tabular}{@{}lcccccc@{}}
\toprule
 & \multicolumn{3}{c}{WER (\%)} & \multicolumn{3}{c}{median (ms)} \\
\cmidrule(lr){2-4}\cmidrule(lr){5-7}
 & $\rc{=}1$ & $\rc{=}2$ & dyn. & $\rc{=}1$ & $\rc{=}2$ & dyn. \\
\midrule
LibriAdapt-US~\cite{mathur2020libriadapt} & 16.7 & 16.6 & 16.3 & 890 & 1608 & \textbf{426} \\
TED-LIUM~3~\cite{hernandez2018tedlium}    & 22.9 & 22.6 & 22.0 & 894 & 1590 & \textbf{506} \\
WSJCAM0~\cite{robinson1995wsjcam0}        & 26.3 & 26.0 & 26.0 & 934 & 1632 & \textbf{574} \\
Common Voice US~\cite{ardila2020common}   & 37.5 & 37.5 & 38.3 & 954 & 1634 & \textbf{704} \\
\bottomrule
\end{tabular}
\end{table}

\begin{table}[b]
\caption{Latency--WER at $\cs{=}16$ on full test-clean ($52.6$k words): per-word
mean, median, and 90th-percentile emission latency (ms) against MFA word ends;
offline is the non-streaming ceiling. Each dynamic $\theta$ is shown as whole-chunk
rollback then $+$\,stable-prefix (same trigger; WER unchanged at this precision), which
cuts the tail. $\theta{=}0.70$ matches the best static accuracy ($6.5\%$) at a fraction
of its latency; $\theta{=}0.85$ gives the lowest median ($268$\,ms) at $7.0\%$ WER.
The confidence-trigger rows use the same strategy~D; SR-CEM pushed further reaches the
lowest WER ($6.3\%$) at higher latency. Random and oracle are the control triggers of
Section~\ref{sec:exp}; the oracle is not deployable.}
\label{tab:dynamic}
\centering
\setlength{\tabcolsep}{3pt}
\begin{tabular}{lcccc}
\toprule
system & WER (\%) & mean & median & $p_{90}$ \\
\midrule
Offline (ceiling)         & 4.5 & --   & --   & --   \\
Static $\rc{=}0$ (causal) & 8.5 & 240  & 238  & 696  \\
Static $\rc{=}1$          & 6.7 & 813  & 860  & 1298 \\
Static $\rc{=}4$          & 6.5 & 2490 & 2910 & 3518 \\
\addlinespace
Dynamic $\theta{=}0.70$ (rollback) & 6.5 & 906 & 442 & 3174 \\
\quad $+$\,stable-prefix           & 6.5 & 630 & 306 & 2800 \\
\quad $+$\,budget $B{=}1$          & 6.8 & 362 & 328 & \textbf{964} \\
Dynamic $\theta{=}0.85$ (rollback) & 7.0 & 551 & 324 & 1496 \\
\quad $+$\,stable-prefix           & 7.0 & 416 & \textbf{268} & 924 \\
\addlinespace
SR-CEM $\theta{=}0.85$ $+$\,stable-prefix & 6.5 & 644 & 312 & 2820 \\
SR-CEM $\theta{=}0.95$ $+$\,stable-prefix & \textbf{6.3} & 919 & 386 & 3164 \\
Top-1 $\theta{=}0.95$ $+$\,stable-prefix  & 6.5 & 688 & 320 & 2920 \\
\addlinespace
Random trigger ($34\%$ deferred) & 7.0 & 1410 & 1078 & 3248 \\
Oracle trigger (rollback) & 6.3 & 696 & 292 & 3072 \\
\addlinespace
Blockwise~\cite{tsunoo2021streaming}      & 7.4 & 1366 & 1270 & 1654 \\
FastEmit~\cite{yu2021fastemit}  & 6.7 & 327  & 326  & 656  \\
\bottomrule
\end{tabular}
\end{table}

\begin{table}[b]
\caption{Generalization to the \textsc{-other} splits at $\cs{=}16$ (out of domain for the
LibriSpeech-360 model); latency uses the same MFA anchoring with per-split references. The dynamic trigger (learned wait-policy, $\theta{=}0.70$,
commit-stable-prefix) matches static $\rc{=}2$ in WER, slightly lower on test-other, at
roughly a third of the median latency. The unbudgeted deferral tail exceeds static
$\rc{=}2$ ($p_{90}$), as on test-clean (Table~\ref{tab:dynamic}, budget row).}
\label{tab:splits}
\centering
\setlength{\tabcolsep}{5pt}
\begin{tabular}{lcccc}
\toprule
split \& system & WER (\%) & mean & median & $p_{90}$ \\
\midrule
\multicolumn{5}{l}{\textit{test-other}}\\
\quad static $\rc{=}2$ & 17.7 & 1405 & 1554 & 2050 \\
\quad dynamic          & 17.4 & 1045 & \textbf{464} & 3216 \\
\multicolumn{5}{l}{\textit{dev-other}}\\
\quad static $\rc{=}2$ & 17.5 & 1396 & 1552 & 2052 \\
\quad dynamic          & 17.4 & 1025 & \textbf{462} & 3220 \\
\bottomrule
\end{tabular}
\end{table}

\subsection{Streaming Baselines}
We evaluate two published streaming systems on the same full test-clean set. The
Offline row in Table~\ref{tab:dynamic} ($4.5\%$) is the non-streaming
ceiling: the ${\sim}2$ percentage-point gap to any streaming row below is the
structural cost of incremental emission, shared by every streaming system here. Both systems use
LibriSpeech train-clean-360 data and $5000$-token BPE vocabulary, a comparable
$12$-layer Conformer encoder, beam-$20$ decoding, and the same per-word latency
measured against MFA word ends.
The blockwise Conformer~\cite{tsunoo2021streaming} (block $40$,
hop $16$, look-ahead $16$) is fixed-latency by design: block geometry is baked into
the encoder and there is no per-token defer; per-token deferral also needs a bounded
encoder, since with a growing receptive field one added chunk shifts representations
arbitrarily far back, leaving no local unit to re-encode. Varying its look-ahead from
$0$ to $24$ gives no better point, up to $15.7\%$ at $1628$\,ms, so we report its best,
$7.4\%$ WER at $1366$\,ms mean per word; our dynamic trigger reaches $6.5\%$ at
$630$/$306$\,ms, lower on both axes. The FastEmit chunk-causal
Conformer--Transducer~\cite{yu2021fastemit} is a frame-synchronous transducer from a
different model family; we include it as a reference point, not a comparable baseline,
because it never defers and its emission timing is fixed at training. It traces its own
curve ($8.2\%$ at $141$\,ms to $5.6\%$ at $710$\,ms mean) and at comparable WER reaches lower
mean latency ($327$ vs $630$\,ms at $6.7$ vs $6.5\%$), as frame-synchronous transducers
do; we do not claim to close that gap.
Our contribution fills the corresponding gap for attention-based models: per-token,
inference-time deferral that re-encodes and re-decodes the deferred chunk, on a
chunk-causal encoder established in prior
work~\cite{zhang2020unified,shi2021emformer} and distinct from our bounded-lookahead
encoder (Section~\ref{sec:encoder}).

\section{Conclusion and Further Research}\label{sec:concl}
We presented two complementary mechanisms for the latency--accuracy trade-off in
streaming attention-based ASR: a \emph{bounded-lookahead chunk encoder} that uses
future-chunk context without a receptive field that grows with the number of layers, and \emph{dynamic
future-chunk decoding} that waits only when a trigger fires; most tokens commit with no
added delay. It matches fixed-lookahead accuracy
at substantially lower median latency ($860\!\to\!306$\,ms at matched WER), with a wait
budget to bound the heavier deferral tail, for attention-based models rather
than frame-synchronous transducers.

The dynamic system is built to match fixed-lookahead WER; the
contribution is \emph{where} latency is spent. The latency reductions hold across the
LibriSpeech splits and on LibriAdapt, TED-LIUM~3, and WSJCAM0, up to $26\%$ WER. On Common
Voice the trigger's latency tracks the high WER, deferring more without lowering it;
the residual error is the domain shift, which lookahead does not fix.

Age-versioning adds training memory and $\rc{+}1$ encoder passes per chunk. A trigger
acts only on \emph{present} uncertainty, and some tokens have none: homophones,
function-word swaps, or tokens a downstream word overturns. These are linguistic errors
that no commit-time acoustic signal can flag; a preliminary language-model signal
ranked deferrals well below acoustic confidence. Future work targets WER monotonicity
in lookahead and a semantic signal that can catch these cases before commitment.

\section*{Acknowledgment}
This work was supported in part by the Flemish Government under the FWO-SBO under Grant
S004923N, and in part by KU Leuven under Grant C24M/22/025. The authors used Claude
(Anthropic) to assist with scripts, with editing the text, and with checking results; all
content was verified by the authors, who take full responsibility for it.


\clearpage


\begin{thebibliography}{00}
\bibitem{watanabe2017hybrid} S.~Watanabe, T.~Hori, S.~Kim, J.~R.~Hershey, and T.~Hayashi, ``Hybrid CTC/attention architecture for end-to-end speech recognition,'' \emph{IEEE J. Sel. Topics Signal Process.}, vol.~11, no.~8, pp.~1240--1253, 2017.
\bibitem{shangguan2021dissecting} Y.~Shangguan, R.~Prabhavalkar, H.~Su, J.~Mahadeokar, Y.~Shi, J.~Zhou, C.~Wu, D.~Le, O.~Kalinli, C.~Fuegen, and M.~L.~Seltzer, ``Dissecting user-perceived latency of on-device E2E speech recognition,'' in \emph{Proc. Interspeech}, 2021.
\bibitem{yu2021fastemit} J.~Yu, C.-C.~Chiu, B.~Li, S.-Y.~Chang, T.~N.~Sainath, Y.~He, A.~Narayanan, W.~Han, A.~Gulati, Y.~Wu, and R.~Pang, ``FastEmit: low-latency streaming ASR with sequence-level emission regularization,'' in \emph{Proc. ICASSP}, 2021.
\bibitem{graves2006ctc} A.~Graves, S.~Fern\'andez, F.~Gomez, and J.~Schmidhuber, ``Connectionist temporal classification: labelling unsegmented sequence data with recurrent neural networks,'' in \emph{Proc. ICML}, 2006, pp.~369--376.
\bibitem{graves2013speech} A.~Graves, A.~Mohamed, and G.~Hinton, ``Speech recognition with deep recurrent neural networks,'' in \emph{Proc. ICASSP}, 2013.
\bibitem{graves2012transduction} A.~Graves, ``Sequence transduction with recurrent neural networks,'' in \emph{Proc. ICML Workshop on Representation Learning}, 2012.
\bibitem{rao2017exploring} K.~Rao, H.~Sak, and R.~Prabhavalkar, ``Exploring architectures, data and units for streaming end-to-end speech recognition with RNN-transducer,'' in \emph{Proc. ASRU}, 2017.
\bibitem{he2019streaming} Y.~He \emph{et al.}, ``Streaming end-to-end speech recognition for mobile devices,'' in \emph{Proc. ICASSP}, 2019.
\bibitem{sainath2019twopass} T.~N. Sainath \emph{et al.}, ``Two-pass end-to-end speech recognition,'' in \emph{Proc. Interspeech}, 2019.
\bibitem{chan2016listen} W.~Chan, N.~Jaitly, Q.~Le, and O.~Vinyals, ``Listen, attend and spell: a neural network for large vocabulary conversational speech recognition,'' in \emph{Proc. ICASSP}, 2016.
\bibitem{tsunoo2019contextual} E.~Tsunoo, Y.~Kashiwagi, T.~Kumakura, and S.~Watanabe, ``Transformer ASR with contextual block processing,'' arXiv:1910.07204, 2019.
\bibitem{zeineldeen2024chunked} M.~Zeineldeen, A.~Zeyer, R.~Schl\"uter, and H.~Ney, ``Chunked attention-based encoder-decoder model for streaming speech recognition,'' in \emph{Proc. ICASSP}, 2024.
\bibitem{zhang2020unified} B.~Zhang, D.~Wu, Z.~Yao, X.~Wang, F.~Yu, C.~Yang, L.~Guo, Y.~Hu, L.~Xie, and X.~Lei, ``Unified streaming and non-streaming two-pass end-to-end model for speech recognition,'' arXiv:2012.05481, 2021.
\bibitem{yao2021wenet} Z.~Yao, D.~Wu, X.~Wang, B.~Zhang, F.~Yu, C.~Yang, Z.~Peng, X.~Chen, L.~Xie, and X.~Lei, ``WeNet: production oriented streaming and non-streaming end-to-end speech recognition toolkit,'' in \emph{Proc. Interspeech}, 2021.
\bibitem{shi2021emformer} Y.~Shi, Y.~Wang, C.~Wu, C.-F.~Yeh, J.~Chan, F.~Zhang, D.~Le, and M.~L.~Seltzer, ``Emformer: efficient memory transformer based acoustic model for low latency streaming speech recognition,'' in \emph{Proc. ICASSP}, 2021.
\bibitem{an2022cuside} K.~An, H.~Zheng, Z.~Ou, H.~Xiang, K.~Ding, and G.~Wan, ``CUSIDE: chunking, simulating future context and decoding for streaming ASR,'' in \emph{Proc. Interspeech}, 2022.
\bibitem{le2024drc} K.~Le and D.~Chau, ``Improving streaming speech recognition with time-shifted contextual attention and dynamic right context masking,'' in \emph{Proc. Interspeech}, 2024.
\bibitem{xia2025mfla} Y.~Xia, H.~Li, C.~Le, M.~Wang, Y.~Sun, X.~Ma, and Y.~Qian, ``MFLA: monotonic finite look-ahead attention for streaming speech recognition,'' in \emph{Proc. Interspeech}, 2025.
\bibitem{moritz2021dual} N.~Moritz, T.~Hori, and J.~Le~Roux, ``Dual causal/non-causal self-attention for streaming end-to-end speech recognition,'' arXiv:2107.01269, 2021.
\bibitem{song2023trimtail} X.~Song, D.~Wu, Z.~Wu, B.~Zhang, Y.~Zhang, Z.~Peng, W.~Li, F.~Pan, and C.~Zhu, ``TrimTail: low-latency streaming ASR with simple but effective spectrogram-level length penalty,'' in \emph{Proc. ICASSP}, 2023.
\bibitem{kang2023delay} W.~Kang, Z.~Yao, F.~Kuang, L.~Guo, X.~Yang, L.~Lin, P.~{\.Z}elasko, and D.~Povey, ``Delay-penalized transducer for low-latency streaming ASR,'' in \emph{Proc. ICASSP}, 2023.
\bibitem{moritz2019triggered} N.~Moritz, T.~Hori, and J.~Le~Roux, ``Triggered attention for end-to-end speech recognition,'' in \emph{Proc. ICASSP}, 2019.
\bibitem{strimel2023ancat} G.~P.~Strimel, Y.~Xie, B.~King, M.~Radfar, A.~Rastrow, and A.~Mouchtaris, ``Lookahead when it matters: adaptive non-causal transformers for streaming neural transducers,'' in \emph{Proc. ICML}, 2023.
\bibitem{huang2020asynchronous} M.~Huang, M.~Cai, J.~Zhang, Y.~Zhang, Y.~You, Y.~He, and Z.~Ma, ``Dynamic latency speech recognition with asynchronous revision,'' arXiv:2011.01570, 2020.
\bibitem{wang2020scout} C.~Wang, Y.~Wu, L.~Lu, S.~Liu, J.~Li, G.~Ye, and M.~Zhou, ``Low latency end-to-end streaming speech recognition with a Scout network,'' in \emph{Proc. Interspeech}, 2020, pp.~2112--2116.
\bibitem{chiu2018mocha} C.-C.~Chiu and C.~Raffel, ``Monotonic chunkwise attention,'' in \emph{Proc. ICLR}, 2018.
\bibitem{gulati2020conformer} A.~Gulati, J.~Qin, C.-C.~Chiu, N.~Parmar, Y.~Zhang, J.~Yu, W.~Han, S.~Wang, Z.~Zhang, Y.~Wu, and R.~Pang, ``Conformer: convolution-augmented Transformer for speech recognition,'' in \emph{Proc. Interspeech}, 2020, pp.~5036--5040.
\bibitem{li2023dynamic} X.~Li, G.~Huybrechts, S.~Ronanki, J.~Farris, and S.~Bodapati, ``Dynamic chunk convolution for unified streaming and non-streaming Conformer ASR,'' in \emph{Proc. ICASSP}, 2023.
\bibitem{dong2024flex} J.~Dong, B.~Feng, D.~Guessous, Y.~Liang, and H.~He, ``Flex attention: a programming model for generating optimized attention kernels,'' arXiv:2412.05496, 2024.
\bibitem{dao2022flashattention} T.~Dao, D.~Y.~Fu, S.~Ermon, A.~Rudra, and C.~R\'e, ``FlashAttention: fast and memory-efficient exact attention with IO-awareness,'' in \emph{Proc. NeurIPS}, 2022.
\bibitem{tsunoo2021streaming} E.~Tsunoo, Y.~Kashiwagi, and S.~Watanabe, ``Streaming Transformer ASR with blockwise synchronous beam search,'' in \emph{Proc. IEEE SLT}, 2021.
\bibitem{mcauliffe2017montreal} M.~McAuliffe, M.~Socolof, S.~Mihuc, M.~Wagner, and M.~Sonderegger, ``Montreal Forced Aligner: trainable text-speech alignment using Kaldi,'' in \emph{Proc. Interspeech}, 2017, pp.~498--502.
\bibitem{liu2020lowlatency} D.~Liu, G.~Spanakis, and J.~Niehues, ``Low-latency sequence-to-sequence speech recognition and translation by partial hypothesis selection,'' in \emph{Proc. Interspeech}, 2020.
\bibitem{polak2022cuni} P.~Pol\'ak, N.-Q.~Pham, T.~N.~Nguyen, D.~Liu, C.~Mullov, J.~Niehues, O.~Bojar, and A.~Waibel, ``CUNI-KIT system for simultaneous speech translation task at IWSLT 2022,'' in \emph{Proc. IWSLT}, 2022.
\bibitem{li2021confidence} Q.~Li \emph{et al.}, ``Confidence estimation for attention-based sequence-to-sequence models for speech recognition,'' in \emph{Proc. ICASSP}, 2021.
\bibitem{qiu2021learning} D.~Qiu \emph{et al.}, ``Learning word-level confidence for subword end-to-end ASR,'' in \emph{Proc. ICASSP}, 2021.
\bibitem{guo2017} C.~Guo, G.~Pleiss, Y.~Sun, and K.~Q. Weinberger, ``On calibration of modern neural networks,'' in \emph{Proc. ICML}, 2017.
\bibitem{jia2026leveraging} Y.~Jia and H.~Van hamme, ``Leveraging beam search information for confidence estimation in E2E ASR,'' \emph{IEEE Open J. Signal Process.}, 2026.
\bibitem{sato2026uncertainty} H.~Sato, A.~Sakuma, R.~Sugano, T.~Kumano, Y.~Kawai, S.~Watanabe, and T.~Ogawa, ``Uncertainty-based streaming ASR with evidential deep learning,'' \emph{IEEE Open J. Signal Process.}, 2026.
\bibitem{selfridge2011stability} E.~Selfridge, I.~Arizmendi, P.~A.~Heeman, and J.~D.~Williams, ``Stability and accuracy in incremental speech recognition,'' in \emph{Proc. SIGDIAL}, 2011.
\bibitem{mcgraw2012estimating} I.~McGraw and A.~Gruenstein, ``Estimating word-stability during incremental speech recognition,'' in \emph{Proc. Interspeech}, 2012.
\bibitem{su2021roformer} J.~Su, Y.~Lu, S.~Pan, A.~Murtadha, B.~Wen, and Y.~Liu, ``RoFormer: enhanced Transformer with rotary position embedding,'' arXiv:2104.09864, 2021.
\bibitem{watanabe2018espnet} S.~Watanabe, T.~Hori, S.~Karita, \emph{et al.}, ``ESPnet: end-to-end speech processing toolkit,'' in \emph{Proc. Interspeech}, 2018.
\bibitem{panayotov2015librispeech} V.~Panayotov, G.~Chen, D.~Povey, and S.~Khudanpur, ``Librispeech: an ASR corpus based on public domain audio books,'' in \emph{Proc. ICASSP}, 2015.
\bibitem{kingma2014adam} D.~P. Kingma and J.~Ba, ``Adam: a method for stochastic optimization,'' in \emph{Proc. ICLR}, 2015.
\bibitem{yu2021dualmode} J.~Yu, W.~Han, A.~Gulati, C.-C.~Chiu, B.~Li, T.~N.~Sainath, Y.~Wu, and R.~Pang, ``Dual-mode ASR: Unify and improve streaming ASR with full-context modeling,'' in \emph{Proc. ICLR}, 2021.
\bibitem{mathur2020libriadapt} A.~Mathur, F.~Kawsar, N.~Bianchi-Berthouze, and N.~D.~Lane, ``Libri-Adapt: a new speech dataset for unsupervised domain adaptation,'' in \emph{Proc. ICASSP}, 2020.
\bibitem{hernandez2018tedlium} F.~Hernandez, V.~Nguyen, S.~Ghannay, N.~Tomashenko, and Y.~Est\`eve, ``TED-LIUM~3: twice as much data and corpus repartition for experiments on speaker adaptation,'' in \emph{Proc. SPECOM}, 2018.
\bibitem{robinson1995wsjcam0} T.~Robinson, J.~Fransen, D.~Pye, J.~Foote, and S.~Renals, ``WSJCAM0: a British English speech corpus for large vocabulary continuous speech recognition,'' in \emph{Proc. ICASSP}, 1995.
\bibitem{ardila2020common} R.~Ardila, M.~Branson, K.~Davis, \emph{et al.}, ``Common Voice: a massively-multilingual speech corpus,'' in \emph{Proc. LREC}, 2020.
\bibitem{vaswani2017attention} A.~Vaswani, N.~Shazeer, N.~Parmar, \emph{et al.}, ``Attention is all you need,'' in \emph{Proc. NeurIPS}, 2017.
\end{thebibliography}
\end{document}